# Investigation of On-Chip Inductors for Fully Integrated DC-DC Converters

Nan Xu, Student Member, IEEE, Wing-Hung Ki, Member, IEEE

Integrated Circuit Design Center, Department of Electronic and Computer Engineering,

The Hong Kong University of Science and Technology, Hong Kong SAR, P. R. China

Email: nxuab@connect.ust.hk, eeki@ust.hk

*Abstract*—On-silicon inductors using a bulk 0.18μm CMOS process have been designed. By shunting different metal layers in parallel, inductor values and quality factors were simulated. Selected inductors were then employed in two open-loop buck converters for comparison: the first used an off-chip discrete diode, and the second used an on-chip active diode. All inductors and converters were sent for fabrication. The fabricated inductors were then measured to have values more than 250 nH over a wide range of frequency, validating the simulation results. The buck converters were switched at 30MHz with a fixed duty ratio of 0.5, to generate an output voltage of 1.2V from an input voltage of 2.4V. The peak efficiency was measured to be 69.1% for a light load current of 12.9 mA.

## I. INTRODUCTION

With the increase in the operating frequency, the footprints of discrete inductors have drastically been decreased, but they still occupy most of the PCB area [1], and system-on-chip (SOC) applications would require passive components such as inductors to be built on-chip [2], for the application of energy harvesting or biomedical scenarios [3]. In [4], an interleaved buck converter switched at 45 MHz with the two inductors of 11 nH each was formed by plating 10μm copper onto the silicon, improving the Q factor and the efficiency reached 64%. In [5], a spiral inductor was designed using bond-wires. In [6], spiral coupled inductors of 2 nH each were designed using metal layers. In [7], the underpass of the spiral on-chip inductor was connected to the output voltage so as to reduce switching loss. In [8], packaging inductors with low DC resistance were designed using a combination of bond-wire and lead-frame. In [9], inductors were formed by the bond-wires connecting the pads and the package pins. Note that post-processing of thick metal layers is expensive while using bond-wire or lead-frame would limit the potential for further minimizing towards SoC. In the case of flip-chip technology, there would even be no bond-wire to use. Therefore, it would be beneficial to investigate using the metal layers of an inexpensive bulk CMOS process to design the power inductor.

The rest of this paper is organized as follows. Section II discusses the modeling of on-chip inductors and the DC-DC converter with an active diode. Section III discusses the design of on-chip inductors and implementing them in two power stages. Section IV shows the simulation and measurement results. Section V presents some concluding remarks and potential improvement.

## II. INDUCTOR MODELING AND THE ACTIVE DIODE

### A. Model review of on-chip inductors

A simple 2-port lumped model was proposed in [10] and is shown in Fig. 2. This model ignores the magnetic coupling between the spiral inductor and the substrate. However, the current of the spiral inductor could induce eddy currents in the underneath substrate with low resistivity, which is in a modern semiconductor process, the resistivity could be as low as 0.01-Ωcm. This will lead to the significant loss of the spiral inductor due to capacitive and inductive coupling, as well as eddy current loss and potential skin effect that may affect the self-resonant frequency of the inductor designed [11].

To account for the above losses, several coupled coils and resistance are added to emulate the substrate ohmic loss, with the primary being the designed spiral inductor $L_{S1}$ and resistance $R_1$, while the secondary having inductance $L_{S2}$ and resistance $R_2$. The corresponding $L_{S3}$ and resistance $R_3$ are from the substrate. Hence, the spiral inductor L is no longer monolithic but couples with $L_{S1}$ and $L_{S2}$. $L_S$, $R_S$, and $C_{S1}$, $C_{S2}$ are lumped for later-on simulation and measurement.

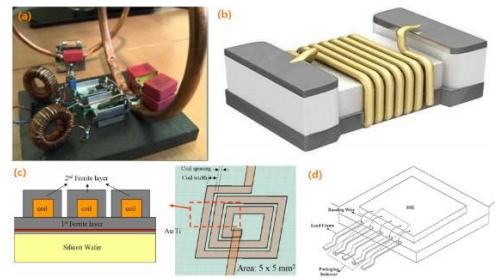

Fig. 1 Diverse inductors (a) Bulky Coil; (b) SMD inductor; (c) Post-processing inductor[12]; (d) Bond-wire and lead-frame inductor[13]

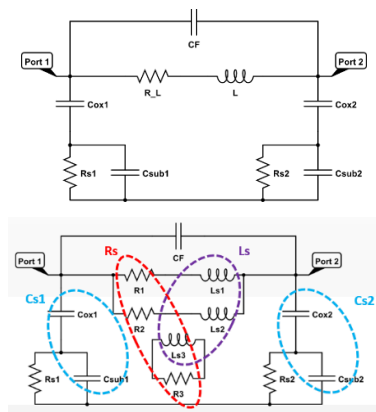

Fig. 2 On-chip inductor: (Top) lumped model; and (Down) improved model

## B. The active diode of the open-loop switching converter

A DC-DC switched-inductor converter consists of a power stage and a feedback control loop [1]. The power stage consists of the inductor, two power switches and the I/O capacitors. The active and the auxiliary power switch is traditionally implemented by a power diode, a passive device. Although the power diode does not need extra control signal, its large voltage drop results in lower efficiency [1]. Hence, for advanced designs, synchronous rectification is employed by replacing the passive diode by an active diode, which consists of a power transistor with low turn-on resistance, and is controlled by a comparator that compares the drain voltage with the source voltage. However, complicate non-overlapping control signals are needed to prevent shoot-through current between the two power transistors, and reverse conduction current should the converter enters discontinuous conduction mode (DCM) in light load condition [2].

Fig. 3 shows a low-loss CMOS active diode proposed for wirelessly powered devices [14]. It utilizes an NMOS that is controlled by a fast voltage comparator to reduce reverse-conduction current. This method can also be used for inductive switching converters to simplify the control mechanism.

## III. PROPOSED ON-CHIP INDUCTOR AND DC-DC POWER STAGES

Investigations on the geometry and layout of on-chip inductors have been conducted [15] [16]. Some of the efforts are summarized in the following subsections.

### A. Design procedure of on-chip inductors

For a CMOS process, metal layers close to the top are thicker, and usually, the closer the thicker; and the smaller the sheet resistance. Higher metal layers are farther away from the substrate and as related capacitive and inductive coupling. One may use the minimum spacing between adjacent turns of a spiral inductor to realize a larger inductance within a smaller silicon area. Even the top metal layers are in fact not very thick (usually only a few microns or less), hence, the induced side-wall capacitance can be neglected.

The width of the metal wire is determined by the current it requires to carry, and the rule of thumb design of current density is 1 mA/μm [1]. Series resistance can be reduced by connecting more than one metal layers in parallel through the adequate number of vias. Moreover, due to coupling, inductance can be enhanced through mutual inductance. The circuit model of two inductors connected in parallel is shown in Fig. 4. The mutual inductance is given by

$$M = k\sqrt{L_1 L_2}, \quad (0 \leqslant k \leqslant 1) \tag{1}$$

The overall equivalent resistance and inductance are

$$R_{eq} = R_1 || R_2 \tag{2}$$

$$L_{eq} = \frac{L_1 \times L_2 - M^2}{L_1 + L_2 - 2M} \tag{3}$$

Therefore, the equivalent quality factor $Q_{eq}$ can be approximated as

$$Q_{eq} = \frac{\omega L_{eq}}{R_{eq}} \approx Q_1 + Q_2 \tag{4}$$

Finally, the self-resonant frequency is:

$$f_{res} = \frac{1}{2\pi\sqrt{L_{eq} \times C}} \tag{5}$$

The halo design can help to balance the inductance and the equivalent series resistance (ESR). With reference to the design of RF inductors, it is recommended that injecting ground isolation layers, such as the patterned ground shield (PGS), between the metal wire and the substrate [17], as well as forming the reverse-biased PN junction on the substrate, could further help to prevent loss and enhance the Q factors.

All the above guidelines are verified with simulation using ASITIC and Sonnet. Fig. 5 shows the optimized inductor that we have designed using Cadence Virtuoso for a DC-DC converter with the following specifications: switching frequency $f_s$ = 30 MHz; $V_{in}$ = 2.4 V, $V_{out}$ = 1.2 V and $I_o$ = 30 mA. The inductor has to fit in a maximum area of 2 mm × 2 mm. The inductor will be fabricated along with the power stages of two open-loop DC-DC converters which are shown in Fig. 6 one using an external passive diode, and another using an on-chip active diode. The parameters are summarized in Table 1.

### B. Power stages of the DC-DC converter

Besides characterizing the designed on-chip inductors, we would also like to verify if the inductors could be used in DC-DC converters, and we decided to compare the performance of the converters using passive diode vs using an active diode. Fig. 7 shows the circuit implementation of the active diode. It consists of an NMOS power switch that is controlled by a comparator realized by a differential common-gate amplifier. The principle of operation is as follows. If the anode (source node) and the cathode (drain node) are at the same potential, the comparator is balanced and both current branches will be 5μA. When the cathode voltage is higher than the anode voltage, the current flows through the transistor M5 will be smaller than that of M6, and the voltage of X1 will be larger than that of X2 and eventually turns off the NMOS power transistor, to prevent reverse current from going through. If the opposite condition occurs, the NMOS power transistor will be turned on. Simulation results of the two converters will be discussed in Sec. IV.

## IV. EXPERIMENTAL RESULTS

On-chip spiral inductors with the two types of power stages were sent for fabrication using a GlobalFoundries 0.18μm process. As shown in Fig. 8, each octagonal inductor occupies 4mm$^2$ area while the power stages are smaller than 1mm$^2$ located at one of the corners. This is a compact floorplan that brings the power stages close to the inductors to minimize conduction loss through metal connecting wires.

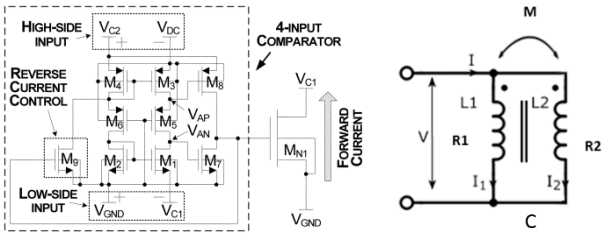

Fig. 3 Active rectifier deployed with reverse current control [14]
Fig. 4 Model of shunt inductors (R$_1$ and R$_2$ are DC resistance of L$_1$ and L$_2$)

The pads labeled "Vx" and "Vout" can be used to characterize the inductor individually without the power stage.

### A. Setups for device and circuit evaluation

The measurement setup of the fabricated on-chip inductors is shown in Fig. 8. The two designs of buck converter were fabricated on the same die. For measurement, the power supplies for the converters and related control signals were externally provided. LCR meters were used for frequency below 2 MHz, and a Vector Network Analyzer was used for frequency running from 1 kHz to 50 MHz.

### B. Characterization of on-chip inductors

The on-chip inductors were measured from 1 kHz to 50 MHz. The parasitic capacitors $C_{ox1}$ and $C_{sub1}$ are merged as $C_{s1}$, and $C_{ox2}$ and $C_{sub2}$ are merged as $C_{s2}$; while $R_{s1}$ is $R_{sub1}$, and $R_{s2}$ is $R_{sub2}$. Fig. 8 shows the measured inductance: the value was around 260 nH from 5 MHz up to 50 MHz, and peaked at 269 nH at 35 MHz. Measured values of $C_{s1}$, $C_{s2}$, $R_{s1}$, and $R_{s2}$ with respect to frequency were also shown. $C_{s1}$ and $C_{s2}$ with respect to frequency followed the same trend; while $R_{s1}$ and $R_{s2}$ with respect to frequency have opposite trends: when one increased the other decreased.

The resonant frequencies $f_{res}$ and the quality factors $Q$ were computed using the measured $L$ and $C$ values. All $f_{res}$ are above 105 MHz, which is around three times of the switching frequency of 30 MHz, and hence the inductors can be used in the buck converters. Moreover, the computed Q increases roughly linearly to around 4 at 30 MHz. Hence, switching at 30 MHz could be lossy but still operative.

TABLE I. PARAMETERS OF THE ON-CHIP INDUCTOR

| Metal width | Spacing | Inner/Outer size | Turns | Top/under layers |
|---|---|---|---|---|
| 50 μm | 1.5 μm | 363μm/1802μm | 14 | M(6//5) / M(4//3) |

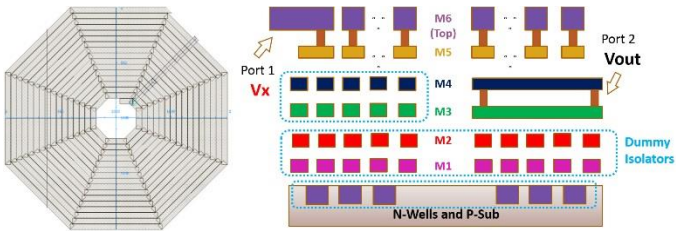

Fig. 5 The optimized on-chip inductor

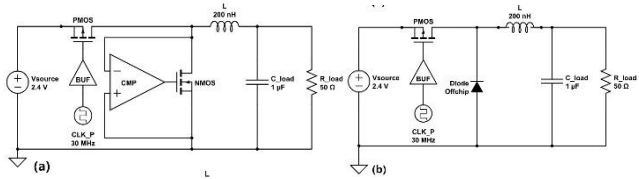

Fig. 6 The power stages with active diode and discrete one

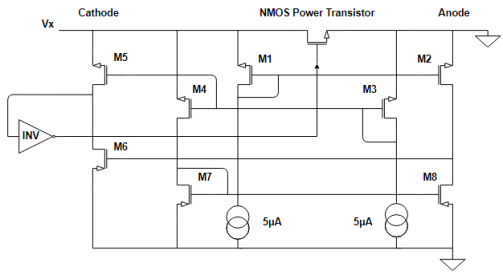

Fig. 7 The configuration of the active diode for NMOS power transistor

### C. Post layout simulation results of power stages

Fig. 10 shows the layouts of the two power stages. After measuring the related parameters (L, $C_{s1}$, $R_{s1}$, etc.) of the on-chip inductors, those values were then written into the Cadence models to give more reliable post-layout simulation results.

The converter with the off-chip passive diode was measured to deliver a larger output current (more than 36 mA), and the converter with the on-chip active diode gave a more stable output voltage, as well as higher efficiency (6% higher). The performance of both converters was compared with an equivalent low dropout regulator (LDR) in Fig. 10. Both converters achieved better efficiency than the LDR, especially when the load resistance was larger than 50 Ω (load current lower than 24 mA). Note that as both converters are open-loop controlled, the output voltage would change when the converters enter DCM.

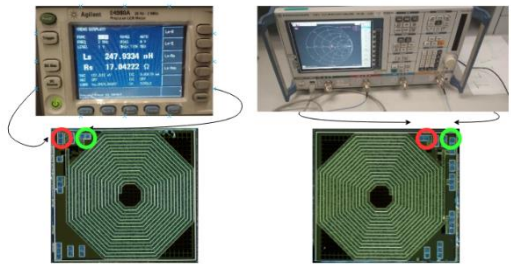

Fig. 8. Measurement of the fabricated inductors with LCR meter (left) and VNA (Right). The Red and Green are "Vx" and "Vout" pads respectively

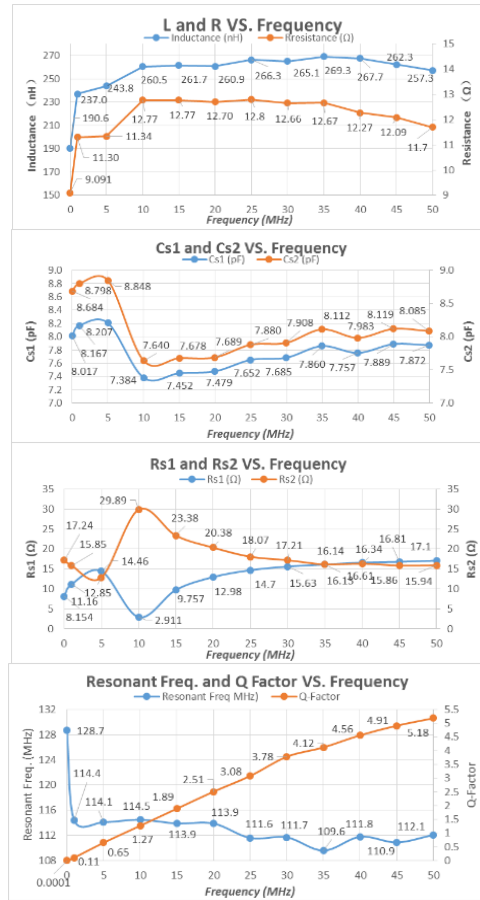

Fig. 9. The measured inductor parameters: L, R, Cs, Rs, Resonant Frequency and Q factor w.r.t.testing frequencies

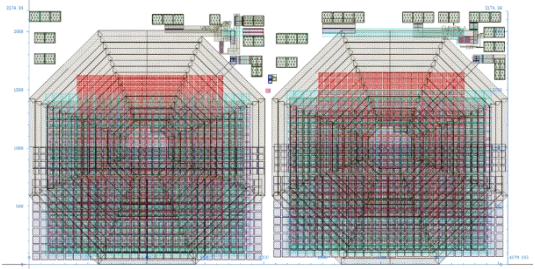

Fig. 10. The layout constructed for post-simulation (Left: with discrete diode. Right: with an active diode)

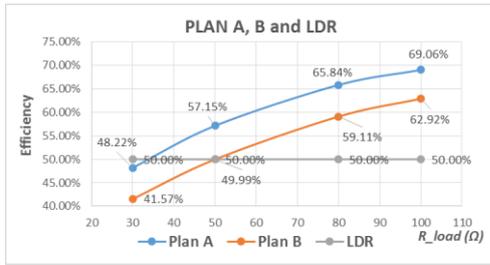

Fig. 11. Comparison of Active (Plan A), Discrete (Plan B) diodes and LDR

TABLE II Post-simulation of converters with active diodes

| $R_{load}$ | $I_{load}$ | $V_{out}$ | $I_{ind}$ | $I_g$ | $\eta$ |
|---|---|---|---|---|---|
| 30Ω | 26.89mA | 0.807V | 26.90mA | 18.75mA | 48.22% |
| 50Ω | 19.28mA | 0.964V | 19.28mA | 13.55mA | 57.15% |
| 80Ω | 15.04mA | 1.203V | 15.04mA | 11.45mA | 65.84% |
| 100Ω | 12.94mA | 1.295V | 12.94mA | 10.11mA | 69.06% |

TABLE III Post-simulation of converters with discrete diodes

| $R_{load}$ | $I_{load}$ | $V_{out}$ | $I_{ind}$ | $I_g$ | $\eta$ |
|---|---|---|---|---|---|
| 30Ω | 36.33mA | 1.091V | 35.34mA | 36.73mA | 44.92% |
| 50Ω | 24.31mA | 1.223V | 24.31mA | 24.72mA | 49.99% |
| 80Ω | 18.12mA | 1.452V | 17.63mA | 18.52mA | 59.11% |
| 100Ω | 15.51mA | 1.550V | 15.52mA | 15.92mA | 62.92% |

TABLE IV Inductor integrated converter comparison

| | *ESSCIRC* [7] | *JSSC* [9] | *JoS* [13] | **This Work** |
|---|---|---|---|---|
| Process | 0.13μm | 0.13μm | 0.18μm | **0.18μm** |
| Inductor Category | Fully on-chip | Bond-wire | Packaging | **Fully On-chip** |
| Vin/out | 3.3V/ 2.5-1.8V | 1.2V/ 0.9V | 1.2-2.4V/ 0.6-1.2V | **2.4V/ 1.2V** |
| I_load | 150 mA | 225 mA | 10 mA | **12.9 mA** |
| η_max | 70.4% | 84.7% | 71.0% | **69.1%** |
| EEF | 22.6% | 11.5% | 29.6% | **27.6%** |
| Freq. | 250 MHz | 100 MHz | 50 MHz | **35 MHz** |
| L | 10.5 nH | 8.5 nH | 20 nH | **240 nH** |
| Q-factor | 14 | NA | NA | **4.12** |
| Area | 4 mm² | 2.25 mm² | 1.68 mm² | **4 mm²** |

## V. CONCLUSIONS

In this work, design of on-chip spiral inductors has been investigated, and they were used in two open-loop buck converters, one with an off-chip passive diode, the other with an on-chip active diode. The inductors and converters were fabricated and measured. The on-chip spiral inductors measured to have an inductance of higher than 240 nH over a wide frequency range. The buck converters have good efficiency, with the peak efficiency of 69.1% for a light load condition of 12.9 mA at 30 MHz.

The performance of the two converters is compared with published works in Table IV. Efficiency Enhancement Factor (EEF) is presented as the figure of merit that indicates the advantage of the converter over that of a linear regular with the same voltage conversion ratio. This work achieved the highest on-chip inductance and relatively high EEF.

REFERENCES

[1] R. W. Erickson, & D. Maksimovic, *Fundamentals of power electronics*. Springer Science & Business Media, 2007.

[2] A. Pressman, *Switching power supply design*. McGraw-Hill, Inc., 1997.

[3] H. Shao, C. Y. Tsui, W. H. Ki, "A micro power management system and maximum output power control for solar energy harvesting applications," *Int. Symp. on Low Power Elec. and Design*, pp. 298-303, 2007.

[4] S. Abedinpour, B.Bakkaloglu, & S. Kiaei, "A multistage interleaved synchronous Buck converter with integrated output filter in 0.18μm SiGe process," *IEEE Trans. on Power Electron.*, vol. 22, no. 6, pp. 2164-2175, 2007.

[5] M. Wens, and M. Steyaert, "A fully-integrated 0.18 μm CMOS DC-DC step-down converter, using a bondwire spiral inductor," *In Proc. IEEE Custom Integr. Circuits Conf.*, pp. 17-20, September, 2008.

[6] J. Wibben, and R. Harjani, "A high-efficiency DC–DC converter using 2 nH integrated inductors," *IEEE J. of Solid-State Circuits*, vol. 43, no. 4, pp. 844-854, 2008.

[7] J. Ni, Z. Hong, and B. Y. Liu, "Improved on-chip components for integrated DC-DC converters in 0.13 μm CMOS," *In Proc. IEEE ESSCIRC*, pp. 448-451, September, 2009.

[8] Y. Ahn, H. Nam, and J. Roh, "A 50-MHz fully integrated low-swing buck converter using packaging inductors," *IEEE Trans. on Power Electron.*, vol. 27, no. 10, pp. 4347-4356, 2012.

[9] C. Huang, & P. K. Mok, "An 84.7% efficiency 100-MHz package bondwire-based fully integrated buck converter with precise DCM operation and enhanced light-load efficiency". *IEEE J. of Solid-State Circuits*, vol. 48, no. 11, pp. 2595-2607, 2013.

[10] C. P. Yue, C. Ryu. J. Lau, T. H. Lee, and S. S. Wong, "A physical model for planar spiral inductors on silicon," *In IEEE International Electron Devices Meeting. Technical Digest,* pp. 155-158, December, 1996.

[11] T. Yeung, J. Lau, H. C. Ho, M. C. Poon, "Design considerations for extremely high-Q integrated inductors and their application in CMOS RF power amplifier," *Radio and Wireless Conf.*, pp. 265-268, 1998.

[12] S. Bae, Y. K. Hong, J. J. Lee, J. Jalli, G. S. Abo, A. Lyle, B. C. Choi, and G. W. Donohoe, "High Q Ni-Zn-Cu ferrite inductor for on-chip power module." *IEEE Trans. on Magn.* vol. 45, no. 10, pp. 4773-4776, 2009.

[13] Z. Zhang, X. Wang, W. Yu, Y. Tan, Y. Yang, and G. Xie, "50 MHz dual-mode buck DC-DC converter," *J. of Semiconduct.*, vol. 37, no. 8, p. 085002, 2016.

[14] Y. H. Lam, W. H. Ki, C. Y. Tsui, "Integrated low-loss CMOS active rectifier for wirelessly powered devices*," IEEE Tran. on Circ. and Syst. II,* pp. 1378-1382, 2006.

[15] H. M. Hsu, "Investigation on the layout parameters of the on-chip inductor," *J. of Microelectronics*, vol. 37, no. 8, pp. 800-803, 2006.

[16] H. M. Hsu, "Analytical formula for inductance of metal of various widths in spiral inductors," *IEEE Trans. on Electron Devices,* vol. 51, no. 8, pp. 1343-1346, 2004.

[17] C. P. Yue, S. S. Wong, "On-chip spiral inductors with patterned ground shields for Si-based RF ICs," *IEEE J. of Solid-State Circuits,* vol. 33, no. 5, pp. 743-752, 1998.